\documentclass[sigconf, 10.5pt, letter]{acmart}

\renewcommand\footnotetextcopyrightpermission[1]{} 
\usepackage{comment}
\usepackage{courier}
\usepackage{algorithmic}
\usepackage{graphicx}
\usepackage{xcolor}
\usepackage{enumitem}
\usepackage{hyperref}
\usepackage{textcomp}
\usepackage[T1]{fontenc}
\usepackage{textcomp}
\usepackage{filecontents}
\usepackage{multirow}
\usepackage{listings}
\usepackage{tikz,url}
\usetikzlibrary{arrows.meta}
\usetikzlibrary{positioning}
\usetikzlibrary{shapes}

\usepackage{amsmath}
\usepackage{fancyhdr}
\usepackage[linguistics]{forest}

\definecolor{codegreen}{rgb}{0,0.6,0}
\definecolor{codegray}{rgb}{0.5,0.5,0.5}
\definecolor{codepurple}{rgb}{0.58,0,0.82}
\definecolor{backcolour}{rgb}{0.95,0.95,0.92}

\lstdefinestyle{mystyle}{ 
    backgroundcolor=\color{backcolour},   
    commentstyle=\color{codegreen},
    keywordstyle=\color{magenta},
    numberstyle=\tiny\color{codegray},
    stringstyle=\color{codepurple},
    basicstyle=\ttfamily\footnotesize,
    breakatwhitespace=false,          
    breaklines=true,                 
    captionpos=b,                    
    keepspaces=true,                 
    numbers=left,                    
    numbersep=5pt,                  
    showspaces=false,                
    showstringspaces=false,
    showtabs=false,                  
    tabsize=2
}
\begin{document}

\title{Robust PDF Files Forensics Using Coding Style}

\author{Supriya Adhatarao}

\affiliation{%
  \institution{Univ. Grenoble Alpes, Inria- France}
}
\email{supriya.adhatarao@inria.fr}

\author{C\'edric Lauradoux}
\affiliation{%
\institution{Univ. Grenoble Alpes, Inria- France}
%  \institution{INRIA Grenoble Rhône-Alpes}
%  \city{Montbonnot Saint-Martin}
  %\country{France}
  }
\email{cedric.lauradoux@inria.fr}

\begin{abstract}
Identifying how a file has been created is often interesting in security. It can be used by both attackers and defenders. Attackers can exploit this information to tune their attacks and defenders can understand how a malicious file has been created after an incident. In this work, we want to identify how a PDF file has been created. This problem is important because PDF files are extremely popular: many organizations publish PDF files online and malicious PDF files are commonly used by attackers. 

Our approach to detect which software has been used to produce a PDF file is based on coding style: given patterns that are only created by certain PDF producers. We have analyzed the coding style of 900 PDF files produced using 11 PDF producers on 3 different Operating Systems. We have obtained a set of 192 rules which can be used to identify 11 PDF producers. We have tested our detection tool on 508836 PDF files published on scientific preprints servers. Our tool is able to detect certain producers with an accuracy of 100\%. Its overall detection is still high (74\%). We were able to apply our tool to identify how  online PDF  services work and to spot inconsistency. 
\end{abstract}


\maketitle

\section{Introduction}

The Portable Document Format (PDF) is a very popular file format used online and created by Adobe Systems. To improve the content and the user experience, the format has evolved over time to support more features like security, searchability, description by metadata\ldots It is standardized as an open format ISO since 2008 and the latest version of the standard is PDF 2.0~\cite{Iso32000}. This popularity has two side effects: (i) PDF files are a popular attack vector and (ii) they contain hidden information that can expose authors sensitive data. Hackers are embedding malicious content to penetrate systems. This is possible because the file format is getting more and more complex. Hackers can also use the PDF files published by an organization to tune their attacks. Many works~\cite{Smith2000,Martin2003,Stevens2011,Smutz2012,Garfinkel2014,Carmony2016,Xu2016,Maiorca2019}   have been dedicated to PDF files security and privacy. 

In this paper, we focus on the following forensics problem: \emph{is it possible to determine how a PDF file has been created using the file itself?} A PDF producing tool detector has applications in offensive security and in incident response. In offensive security, it can be used to determine which software is used by an author or by an organization to create and view PDF files. The attackers can find vulnerabilities\footnote{Many
vulnerabilities have been found in the past in PDF viewers: 1090 vulnerabilities have been reported in December 2020 according to \url{https://cve.mitre.org}.} corresponding to the PDF viewer identified. Then, the attacker can craft and send malicious PDF files to the organization thanks to the knowledge obtained from PDF files. 

In incident response, a PDF producing tool detector is valuable to understand how a malicious PDF file has been created. It is a useful step toward an attack attribution. 

The most simple approach to design a PDF producing tool detector consists to look at the file metadata. By default, PDF producer tools put many information in the field \texttt{Creator} and \texttt{Producer} of the file's metadata. It is possible to find the name of the producer tool and its version as well as details on the Operating System. Unfortunately, metadata are not a reliable source of information: they can be easily modified using tools like \texttt{exiftool}\footnote{\url{https://exiftool.org/}} or using sanitization tools like Adobe Acrobat. 

We have designed a robust PDF producing tool detector based on the coding style of the file. The PDF standard~\cite{Iso32000} defines the language that is supported by PDF viewers. Developers of PDF producer tool have their own interpretation of the PDF language. Therefore, it is likely that their coding style is reflected on the output of their PDF producer tool. There are coding style elements~\cite{Kernighan1978} in PDF files which can be used to identify the producer tool.

To observe the coding style of PDF files, we have created a dataset of 900 PDF files using 11 popular PDF producer tools. We have compared the different files to identify the pattern in each section of the PDF files. We created 192 rules in regular expression engine to identify these patterns and detect the PDF producer tool. Then, we have gathered 508836 PDF files downloaded from  scientific preprints to test the efficiency of our detection tool. We are able to detect PDF files created by \texttt{LibreOffice} and \texttt{PDFLaTeX} tool with an accuracy of 100\%. PDF files created by \texttt{Microsoft Office Word} and \texttt{Mac OS X Quartz} were detected with an accuracy greater than 90\%. More generally, it correctly detected the producer tool of 74\% of the PDF files in our dataset. Our study is less conclusive for the detection of the Operating System (only 32\% of correct guesses).   

This paper is organized as follows. In Section~\ref{sec:PDF} we describe the internal structure of a PDF file. In Section~\ref{sec:eco} we describe PDF ecosystem and the tools we have chosen for our analysis. We give examples of coding style and programming patterns found in PDF files in Section~\ref{sec:beyond}. We present and test our detector in Section~\ref{sec:res}. Finally, we position our results to the related work done on PDF files in Section~\ref{sec:related}. Our comparison includes the reverse engineering of arXiv PDF detector tool.\\
\section{Background on PDF Standard}\label{sec:PDF}
%https://nubuntu.org/postscript-vs-pdf

Portable Document Format (PDF) is based on the PostScript language. It has been standardized as ISO 32000 and was developed by Adobe in 1993 to present documents. This document is independent of application software, hardware, and Operating Systems used and hence widely used. Apart from text, a PDF includes information such as fonts, hyperlinks, instructions for printing, images, keywords for search and indexing\ldots Since 2008, PDF is standardized as an open format ISO and the latest version of the standard is PDF 2.0~\cite{Iso32000}. 

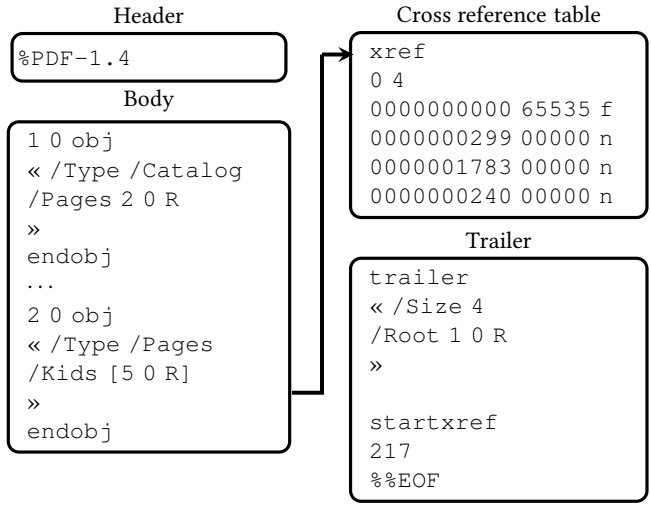
\begin{figure}[ht]
\tikzset{
    state/.style={
           rectangle,
           rounded corners,
           draw=black, very thick,
           minimum height=2em,
           inner sep=2pt,
           text centered,
           },
}
\begin{center}
\begin{tikzpicture}[]

 % Position of QUERY
 % Use previously defined 'state' as layout (see above)
 % use tabular for content to get columns/rows
 % parbox to limit width of the listing
 
% \node[state,text width=3.6cm,align=left,anchor = north] (HEADER)
% \begin{tabular}{l}\texttt{\%PDF-1.4}\end{tabular};
% \node[above=0 of HEADER] (header_title) {Header};

 \node[state,text width=3.5cm,align=left,anchor = north] (HEADER)
 {\texttt{\%PDF-1.4}};
 \node[above=0 of HEADER] (header_title) {Header};

 % State: ACK with different content
 \node[state,    	% layout (defined above)
  text width=3.6cm, 	% max text width
  below=0.5 of HEADER, 	% Position is to the right of QUERY
     anchor = north,
  node distance=0.45cm, 	% distance to QUERY
    align=left] (BODY) 	% posistion relative to the center of the 'box'
 {%
 $\begin{array}{l} 	% content
  \texttt{1 0 obj}\\
  \texttt{<< /Type /Catalog}\\
  \texttt{/Pages 2 0 R}\\
\texttt{>>} \\
\texttt{endobj} \\
\cdots\\
\texttt{2 0 obj}\\
  \texttt{<< /Type /Pages}\\
  \texttt{/Kids [5 0 R]}\\
\texttt{>>} \\
\texttt{endobj} \\
 \end{array}$
 };
 \node[above=0 of BODY] (body_title) {Body};

 % STATE QUERYREP
 \node[state,
  right of=HEADER,
  align=left,
   anchor = north,
   yshift = 0.35cm,
    node distance=4.65cm, 	% distance to QUERY
  text width=3.8cm] (XREF)
 {%
 $\begin{array}{l}
  \texttt{xref}\\
    \texttt{0 4}\\
    \texttt{0000000000 65535 f}\\
    \texttt{0000000299 00000 n}\\
    \texttt{0000001783 00000 n}\\
        \texttt{0000000240 00000 n}\\

 \end{array}$
 };
 \node[above=0 of XREF] (xref_title) {Cross reference table};

 % STATE EPC
 \node[state,
  below=0.5 of XREF,
  node distance=1.45cm,
 align=left,
 anchor = north,
  text width=3.8cm] (Trailer)
 {%
 $\begin{array}{l}
    \texttt{trailer}\\
 \texttt{<< /Size 4}\\
  \texttt{/Root 1 0 R}\\
\texttt{>>} \\
\\
  \texttt{startxref} \\
  \texttt{217} \\
  \texttt{\%\%EOF} \\
 \end{array}$
 };
 \node[above=0 of Trailer] (trailer_title) {Trailer};

  \draw[line width=1.5pt] (1.9,-4.8) -- (2.3,-4.8) -- (2.3,-0.3) ;
 \draw[line width=1.5pt,->,>=stealth] (2.3,-0.3) -- (2.72,-0.3);

\end{tikzpicture}
\caption{\texttt{PDF} in a nutshell.}
\label{fig:format}
\end{center}
\end{figure}

In this section, we will outline the structure of a PDF file and how metadata information is stored. PDF document has a specific file structure, Figure~\ref{fig:format} describes the basic structure~\cite{Warnock2001}. It is organized into four parts: \emph{header}, \emph{body}, \emph{cross reference table} and \emph{trailer}. 
%We briefly explain the structure of each section.
%%%%%%%%%%%%%%%%%%%%%%%%%%%%%%%%%%%%%%%%%%%%%%%%%%%%%%%%%%%%%%%%%%%%%%%
%%%%%%%%%%%%%%%%%%%%%%%%%%%%%%%%%%%%%%%%%%%%%%%%%%%%%%%%%%%%%%%%%%%%%%%

\textbf{Header}: Every \texttt{PDF} document starts with the same magic number \texttt{\%PDF} (\texttt{0x25 50 44 46}) with \texttt{\%} being the comment symbol. The version of specification used to encode the file is then appended: \texttt{-m.n} where \texttt{m} is the major number and \texttt{n} is the minor number (Figure~\ref{fig:format}). The header can be composed of a second comment line if the file contained some binary data which is often the case. It consists of at least four binary characters~\cite{Warnock2001}.\\

\textbf{Body}:  The body is a collection of indirect objects. It is the actual content of the document as viewed by the user. These objects represent the fonts, pages, sampled images and object streams of the PDF document. There are eight objects: boolean, numbers, strings, names, arrays, dictionaries, streams and the null object. All the objects in the document are labeled so that they can be referred to by other objects and the label is unique for each object in the PDF document.\\

\textbf{Cross reference table}: This section consists of a collection of entries which gives the byte offsets for each indirect object in the document. This table is essentially used for quick and random access to objects. Cross reference table starts with the keyword \texttt{xref} hence it is also called \texttt{xref table}.\\

\textbf{Trailer}: The trailer part of the PDF document is used for quick access to find the cross reference table and certain special objects in the document. The last line of a PDF document is an end-of-file marker \texttt{\%\%EOF}. This marker is  preceded by a byte offset and the \texttt{startxref} keyword which indicates the first entry of the cross reference table.

There are several different PDF producer tools to create a PDF document and these PDF producer tools use these 4 sections as the basis for creation of a PDF document. \\
 
\textbf{PDF Metadata}: Apart from the content as viewed by the user, PDF files also include some metadata. PDF metadata stores all the information related to the PDF document.  This metadata information is stored in an object in the body section of the document. It is a special object which can be easily altered or removed without having any influence on the rest of the content of the document. PDF producer tools store metadata either in a \texttt{document information dictionary} or in a \texttt{metadata stream}.

\section{PDF ecosystem}\label{sec:eco}

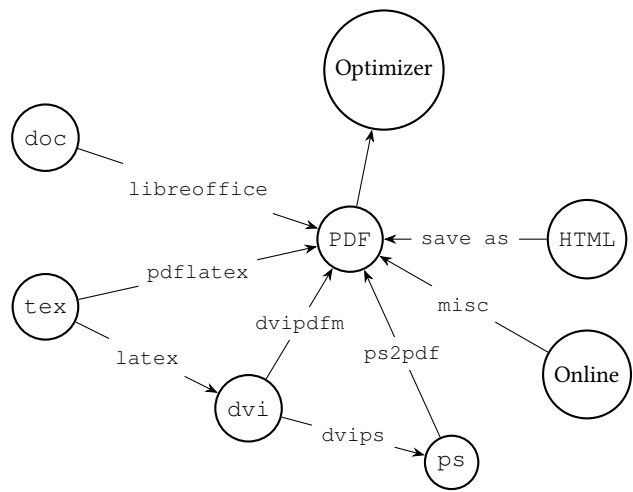
\begin{figure}[ht]
\begin{center}
\begin{tikzpicture}[scale=0.9]
\begin{scope}[every node/.style={circle,thick,draw}]
    \node (doc) at (-2,2.5) {\texttt{doc}};
    \node (tex) at (-2,0) {\texttt{tex}};
        \node (dvi) at (1,-1.5) {\texttt{dvi}};
        \node (ps) at (4,-2.3) { \texttt{ps} };

    \node (Optimizer) at (3,3.5) {Optimizer};
    \node (PDF) at (2.5,1) {\texttt{PDF}};
    \node (Online) at (6,-1) {Online};
    \node (html) at (6,1) {\texttt{HTML}};
\end{scope}

\begin{scope}[>={Stealth[black]},
              every node/.style={fill=white,circle},
              every edge/.style={draw}]
    \path [->] (doc) edge node {\small{\texttt{libreoffice}}} (PDF);
	    \path [->] (tex) edge node {\small{\texttt{pdflatex}}} (PDF);
    	   % \path [->] (Online) edge  (PDF);
    	    %\path [->] (OS) edge  (PDF);
    	    \path [->] (tex) edge node {\small{\texttt{latex}}} (dvi);
    	    \path [->] (dvi) edge  node {\small{\texttt{dvips}}} (ps);
    	    \path [->] (ps) edge node[rectangle] {\small{\texttt{ps2pdf}}} (PDF);
    	    \path [->] (dvi) edge node[rectangle] {\small{\texttt{dvipdfm}}} (PDF);

        	\path [->] (html) edge node {\small{\texttt{save as}}} (PDF);
        	\path [->] (Online) edge node {\small{\texttt{misc}}} (PDF);
			\path [->] (PDF) edge (Optimizer);
   % \path [->] (B) edge  (C);

  %  \path [->] (C) edge node {$5$} (F);
  %  \path [->] (E) edge node {$8$} (F);
  %  \path [->] (B) edge[bend right=60] node {$1$} (E);
\end{scope}
\end{tikzpicture}
\end{center}
\caption{Different options available to create a PDF file.}\label{fig:PDFeco}
\end{figure}

Writing directly PDF commands being difficult for a human, there exist several options to create a PDF file using different file formats ( Figure~\ref{fig:PDFeco}). There is a large ecosystem of PDF creation tools, converters and optimizers which are available online or locally. Many editors also propose in their software the option to convert certain file formats into PDF. Most people are accustomed to convert user-friendly (rich text) documents (doc/docx, ppt) into PDF files. These conversions can be straight-forward, for instance, they are integrated in word processing applications such as \texttt{Microsoft Office} and \texttt{LibreOffice} which transform the doc/docx files into PDF files. Intermediary conversions maybe needed for some file formats, like for the \LaTeX chain \texttt{tex} $\rightarrow$ \texttt{dvi} $\rightarrow$ \texttt{ps} $\rightarrow$ \texttt{PDF}. Popular Operating Systems (OS) and even browsers commonly provide support to print content (html pages, images\ldots) into PDF files. There are several online tools which convert different file formats into PDF files.\\
%---------------------------------------------------------------------------
%---------------------------------------------------------------------------

We broadly classify PDF file producer tools into five categories: OS-based tools, word processors, \LaTeX processors, browsers and optimizer/transformation tools. We expected that each PDF producer tool has its own way to create the PDF code. To verify our guess, we chose 11 PDF producer tools (Table~\ref{tools}) which represent the different categories in Figure~\ref{fig:PDFeco}. We have used these 11 PDF producer tools on three different Operating Systems: Microsoft Windows (Windows 10) , MAC OS (10.15.7) and Linux (Ubuntu 18.04.4 LTS ) Operating  Systems whenever it was possible.

%%%%%%%%%%%%%%%%%%%%%%%%%%%%%%%%%%%%%%%%%%%%%%%%%%%%%%%%%%%%%%%%%%%%%%
\begin{table}
\begin{center}
\begin{tabular}{|l|}
\hline
Producer tools\\  \hline\hline

Acrobat Distiller 		\\
Microsoft Office Word   \\
LibreOffice 			\\
Ghostscript 			\\
Mac OS X Quartz 		\\
PdfTeX 					\\
SKia/PDF 				\\
Cairo 					\\
xdviPDFmx 				\\
LuaTeX 		            \\
PDFLaTeX 				\\ 
\hline
%total & 769&&

\end{tabular}
%\caption{Our methodology results for PDF producer tool detection.}
%\label{14}
\caption{11 PDF producer tools.}
\label{tools}
\end{center}
\end{table}
%------------------------------------------------------------------------%

Our dataset includes 25 source documents for Microsoft Word compatible software and 30 source documents for \LaTeX{} compilation chains. These documents include many different elements like text, images, tables, equations\ldots to be representative of the usual content found in a PDF file. Then, we apply the 11 producer tools on our source documents. It allows us to observe the differences and patterns between the PDF files produced from the same source file.  our goal was to create unique producer signature or fingerprint.

It is important to notice that \texttt{PdfTeX} tool name used in this paper is the label of \LaTeX{} software whereas \texttt{PDFLaTeX} is a standard label used by open archive HAL\footnote{\url{https://hal.archives-ouvertes.fr/?lang=en}} for it's PDF producer tool. We could not create a dataset of PDF files for PDFLaTeX tool. Since we did not want to pollute the website of HAL with our test files, we only downloaded some random PDF files created by HAL for our analysis.

\section{Elements of Coding Style}\label{sec:beyond}
We have analyzed separately the elements of coding style in the four sections of a PDF file: header, body, cross reference table and trailer. 

\subsection{Header}

The header section of PDF files has always the same organization across all the producer tools. It consists of two comments line: first comment line always starts with the same magic number \texttt{\%PDF} (Figure~\ref{fig:format}) and is often followed by another comment if the file contained some binary data. This comment in the second line is left undefined by the specification. During our analysis, we observed that it is often there and producer tools leave different values in the file.  Listing~\ref{header2} shows an header section of the \texttt{Microsoft Office Word} tool. All the PDF files created by this tool include the same binary data as the second comment. 

\begin{center}
\lstset{ %
    basicstyle=\ttfamily\footnotesize,
    frame=single,
    keywordstyle=\color{blue},
    language=Bash,
    showstringspaces=false,
    caption={Binary data - Associated to Microsoft Office Word tool.},
    captionpos=b,
    label=header2,
    morekeywords={blue},
}
\begin{lstlisting}[basicstyle=\small]
%PDF-1.7
%\B5\B5\B5\B5
\end{lstlisting}
\end{center}

%-------------------------------------------------------------------------

Table~\ref{tab:header} shows the different binary data associated to the 11 producer tools. Some values are shared by different tools and some are specific to a distribution and Operating System. For instance, \texttt{OxE2E3CFD3} is the binary value associated to the tool \texttt{Acrobat Distiller} and it is unique. Our analysis  also showed that it is possible that one tool uses several values across different Operating Systems.  In our analysis, LuaTeX uses 2 different values. It depends on the \LaTeX{} distribution and the OS  used. \texttt{OxD0D4C5D8} value is shared by two PDF producer tools \texttt{pdfTeX, LuaTeX}. It is also interesting that this value is specifically associated to Texlive distribution on Linux system only.  This value could be used in the detection of the OS used to create the PDF file.

Nine producer tools in our evaluation have unique values and hence this value can be used to directly reveal the PDF producer tool. This value can be considered as \emph{producer magic number}. Removing the \emph{producer magic number} or even altering it does not have any influence on the display of a PDF file. It is not necessarily a robust method to identify a PDF producer tool because it can be easily modified. Still, the \emph{producer magic number} can significantly help in the detection of the PDF producer tool or at least narrow down the identification to a small set of tools. Table~\ref{tab:header} shows the producer magic number and the associated Operating System, for \LaTeX tools, it also shows the distribution.

%%%%%%%%%%%%%%%%%%%%%%%%%%%%%%%%%%%%%%%%%%%%%%%%%%%%%%%%%%%%%%%%%%%%%%%
\begin{table*}[!ht]
\renewcommand{\arraystretch}{1.2}
\begin{center}
\begin{tabular}{|r|l|l|l|}
\hline
Unique Binary Data & PDF producer tools & OS- & Distribution-\\
(in hexadecimal) & & Windows, Linux, Mac OS X& TeXLive/MikTeX\\ \hline\hline
\texttt{OxE2E3CFD3} & Acrobat Distiller & 3 OSs &-\\
\texttt{OxB5B5B5B5} & Microsoft Office Word & Windows & -\\
\texttt{OxD0D4C5D8} & pdfTeX & 3 OSs & TeXLive and MikTeX\\
\textbf{\texttt{OxD0D4C5D8}} & \textbf{LuaTeX} & \textbf{Linux} & \textbf{TexLive}\\
\textbf{\texttt{OxCCD5C1D4C5D8D0C4C6}} & \textbf{LuaTeX} & \textbf{Linux} & \textbf{MikTeX}\\
\textbf{\texttt{OxCCD5C1D4C5D8D0C4C6}} & \textbf{LuaTeX} & \textbf{Mac OS X and Windows} & \textbf{TeXLive and MikTeX}\\
\texttt{OxE4F0EDF8} & xdvipdfm & 3 OSs & TeXLive and MikTeX\\
\texttt{Oxc7ec8fa2} & GhostScript & 3 OSs & TeXLive and MikTeX\\
%\texttt{BF F7 A2 FE} & qpdf\\
\texttt{Oxc3a4c3bcc3b6c39f} & LibreOffice & Linux & -\\
\texttt{OxC4E5F2E5EBA7F3A0D0C4C6}& Mac Os X & Mac OS X & - \\
\texttt{OxB5EDAEFB} & cairo & 3 OSs & -\\
\texttt{OxD3EBE9E1}& Skia & 3 OSs & -\\
\texttt{OxF6E4FCDF} & PdfLaTeX & online &-\\
\hline
\end{tabular}
\end{center}
\caption{Header- producer magic number and the associated producer tools.}
\label{tab:header}
\end{table*}

\subsection{Body}

%The body part is a huge collection of indirect objects representing the text, fonts, pages, sampled images and object streams of the PDF (Figure~\ref{fig:format}). There are eight objects: boolean, numbers, strings, names, arrays, dictionaries, streams and the null object. %Potential differences across producer tools include the way objects are created, stored, ordered, identified and the contents (metadata information coding, length of the encoded text, images and different fonts). 

The body section of a PDF file is a collection of indirect objects. It is the actual content of the document as viewed by the user. These objects represent the fonts, pages, sampled images and object streams of the PDF file. There are eight types of objects: \textit{boolean, numbers, strings, names, arrays, dictionaries, streams} and \textit{the null object}. All the objects in the document are labeled so that they can be referred to by other objects. The label is unique for each object in the PDF file. Each object is identified by two positive integers, the first number is the object number followed by a second number that is used as a generation number. Initially all the generation numbers are set to zero and can be changed when the document is updated. We observed the following differences between the different producer tools:
\begin{itemize}
\setlength\itemsep{0.4em}
\item Number and type of keys used to describe objects;
\item Use of escape sequences (\textbackslash{}n (newline), \textbackslash{}r (carriage return) and \textbackslash{}t (tab) etc..) in literal strings;
\item Total number of objects created;
\item Arrangement of the objects (random, increasing order, incremental etc..);
\item The way metadata information is stored;
\item Length of the encoded text and images;
\item The way font encoding information is stored.
\end{itemize}

Listing~\ref{body1} and ~\ref{body2} shows example of the same stream object created by \texttt{pdfTeX} and \texttt{LuaTeX} tools. The way they are encoded is different but PDF viewers will display the same output. Even if the object ID and all the keys (Length, Filter, FlateDecode\ldots) used look alike for these two tools, it is possible to distinguish them using the order of arrangement of keys and the use of escape sequences (\textbackslash{}n (newline) and spaces).  Such  differences across different objects encoded in the PDF files can be used in the detection of producer tools. 

\begin{center}
\lstset{ %
    basicstyle=\ttfamily\footnotesize,
    frame=single,
    keywordstyle=\color{blue},
    language=Bash,
    showstringspaces=false,
    caption={Object encoding using pdfTeX tool},
    captionpos=b,
    label=body1,
    morekeywords={blue},
}
\begin{lstlisting}[basicstyle=\small]
4 0 obj
<</Length 2413      /Filter/FlateDecode>>
stream
.....
endstream
endobj
\end{lstlisting}
\end{center}

\begin{center}
\lstset{ %
    basicstyle=\ttfamily\footnotesize,
    frame=single,
    keywordstyle=\color{blue},
    language=Bash,
    showstringspaces=false,
    caption={Object encoding using LuaTeX tool.},
    captionpos=b,
    label=body2,
    morekeywords={blue},
}
\begin{lstlisting}[basicstyle=\small]
4 0 obj
<<
/Length 2006      
/Filter /FlateDecode
>>
stream
.....
endstream
endobj
\end{lstlisting}
\end{center}

%Listing~\ref{microBody} shows an object created by \texttt{Microsoft Office Word} tool and it contains some text data. When we compare this to \texttt{pdfTeX} and \texttt{LuaTeX}

%\begin{center}
%\lstset{ %
%    basicstyle=\ttfamily\footnotesize,
%    frame=single,
%    keywordstyle=\color{blue},
%    language=Bash,
%    showstringspaces=false,
%    caption={text object encoding Microsoft Office Word},
%    captionpos=b,
%    label=microBody,
%    morekeywords={blue},
%}
%\begin{lstlisting}
%4 0 obj
%<</Filter/FlateDecode/Length 4900>>
%stream
%......
%endstream
%endobj
%\end{lstlisting}
%\end{center}

\subsection{Cross reference table}

Cross reference table or the xref table gives the offsets (in bytes) for each indirect object which is used for quick and random access to objects in the body section (Figure~\ref{fig:format}). This section of a PDF file is optional and many producer tools do not include it. Some producer tools use linearization or incremental saves and the information related to this table is encoded in the trailer object. Cross reference table is always coded in the same way across the different tools, when it is present.

Listing~\ref{xref} gives an example of an xref table generated by \texttt{Microsoft Office Word} tool. The cross reference table starts with the keyword \texttt{xref} followed by two numbers separated by a space. The first number indicates the object number of the first object in the subsection.  The second number indicates the total number of entries in the subsection. Listing~\ref{xref} shows an example of the cross reference table extracted from a PDF file created using \texttt{Microsoft Office Word} tool. The table consists of single subsection entry with 5 objects from 0 to 5, where 0 is the first object and 5 is the number of entries in the subsection of the cross reference table. 0 5 also indicates that there are 5 consecutive objects. 

Each cross reference entries (one per line) is associated to exactly one object and it is 20 bytes long and has the format \texttt{"nnnnnnnnnn ggggg n/f eol"}, where the first 10 bytes are \texttt{nnnnnnnnnn}, indicating the byte offset of the referenced entry, followed by a space and then by 5 digits entry \texttt{ggggg}, which is a generation number of the object followed by a space and then a \texttt{n}, where \texttt{n} is a literal keyword to indicate that the object is in use while \texttt{f} is used to indicate that an object is free and this is followed by a space and last 2 bytes constituting the \texttt{end-of-line}.

\begin{center}
\lstset{ %
    basicstyle=\ttfamily\footnotesize,
    frame=single,
    keywordstyle=\color{blue},
    language=Bash,
    showstringspaces=false,
    caption={Cross reference Table - Microsoft Office Word},
    captionpos=b,
    label=xref,
    morekeywords={blue},
}
\begin{minipage}{3.3in}
\begin{lstlisting}[basicstyle=\small]
xref
0 5
0000000010 65535 f
0000000017 00000 n
0000000166 00000 n
0000000222 00000 n
0000000486 00000 n
\end{lstlisting}
\end{minipage}
\end{center}

Since this section of PDF file has same patterns across all the producer tools, it's presence or absence can narrow down the detection of PDF producer tools to a smaller set of candidate tools. 

We found only nine tools among 11 that include the cross reference table:  Acrobat Distiller and xdviPDFmx do not include this table. We observed that \texttt{pdfTeX} tool includes the table only in \texttt{MikTeX} distribution for all three OSs whereas the table has been removed for \texttt{TexLive} distributions. %The presence or absence of this table narrow downs the detection of PDF producer tools to a smaller set of candidate tools. 

\subsection{Trailer}

The trailer part is used for quick access to find the cross reference table and certain special objects in the document. The trailer part is very interesting to detect the producer tool used. The last object present in this part includes \texttt{Root} information and some other keys-values. It is  possible to distinguish producer tools based on the keys used to describe the trailer object.  
 We have chosen two PDF files with same content created using \texttt{LibreOffice} and \texttt{Microsoft Office Word} tools, Listing~\ref{tail1} and ~\ref{tail2} shows example of the trailer objects. 

Both these tools have completely different coding style for the same content of the file and hence can lead to the detection of producer tool. It is interesting to note that \texttt{Microsoft Office Word} tool includes 2 trailer objects which is unique style associated to only this tool.
% different compared to any other tool.
  
\begin{center}
\lstset{ %
    basicstyle=\ttfamily\footnotesize,
    frame=single,
    keywordstyle=\color{blue},
    language=Bash,
    showstringspaces=false,
    caption={Trailer object- Libreoffice},
    captionpos=b,
    label=tail1,
    morekeywords={blue},
}

\begin{minipage}{3.5in}
\begin{lstlisting}[basicstyle=\small]
trailer
<</Size 14/Root 12 0 R
/Info 13 0 R
/ID [ <438A4EF8B552AF586C55DFFE40065998>
<438A4EF8B552AF586C55DFFE40065998> ]
/DocChecksum /7C2B6DC7F4AF6CC658C0703D8002E3D4
>>
\end{lstlisting}
\end{minipage}
\end{center}
 
\begin{center}
\lstset{ %
    basicstyle=\ttfamily\footnotesize,
    frame=single,
    keywordstyle=\color{blue},
    language=Bash,
    showstringspaces=false,
    caption={Trailer object- Microsoft Office Word},
    captionpos=b,
    label=tail2,
    morekeywords={blue},
}
\begin{minipage}{3.5in}
\begin{lstlisting}[basicstyle=\small]
trailer
<</Size 25/Root 1 0 R/Info 9 0 R/ID[<70265267FB5C68469F73B4AB7F5E4003><70265267FB5C68469F73B4AB7F5E4003>] >>
startxref
46566
%%EOF
xref
0 0
trailer
<</Size 25/Root 1 0 R/Info 9 0 R/ID[<70265267FB5C68469F73B4AB7F5E4003><70265267FB5C68469F73B4AB7F5E4003>] /Prev 46566/XRefStm 46274>>
\end{lstlisting}
\end{minipage}
\end{center}

We have listed all the  different keys  in the trailer object in Table~\ref{tab:trailer}. \emph{Tools like LibreOffice, Acrobat Distiller and Microsoft Office Word have their own keys that are distinguishable from any other tools and hence are unique and tool specific}. \texttt{LuaTeX, pdfTeX, Ghostscript} and \texttt{Mac OS X Quartz} share the same set of keys, but the order of arrangement of these keys is different and hence potentially lead to the detection of a producer tool.

%%%%%%%%%%%%%%%%%%%%%%%%%%%%%%%%%%%%%%%%%%%%%%%%%%%%%%%%%%%%%%%%%%%%%
\begin{table*}[!ht]
\renewcommand{\arraystretch}{1.2}
\begin{center}
\begin{tabular}{|r|l|}
\hline
Producer Tool  & Key strings in Trailer section\\ \hline\hline
Acrobat Distiller& /DecodeParms /Columns /Predictor /Filter /FlateDecode /ID /Info /Length /Root /Size /Type /XRef /W\\
%qpdf&/Type /XRef /Length /Filter /FlateDecode /DecodeParms /Columns /Predictor /W /Size /ID \\
TexLive LuaTeX  &/Type /XRef /Index /Size /W /Root /Info /ID /Length /Filter /FlateDecode\\
TexLive pdfTeX  & /Type /XRef /Index /Size /W /Root /Info /ID /Length /Filter /FlateDecode\\
MikTeX LuaTeX&trailer /Size /Root /Info /ID\\
MikTeX pdfTeX&trailer /Size /Root /Info /ID \\
Ghostscript&trailer /Size /Root /Info /ID \\
xdvipdfm&/Type /XRef /Root /Info /ID /Size /W /Filter /FlateDecode /Length\\
Microsoft Office Word&trailer /Size /Root /Info /ID /Prev /XRefStm \\
LibreOffice&trailer /Size /Root /Info /ID /DocChecksum \\
Mac OS X Quartz&trailer /Size /Root /Info /ID\\
cairo&trailer /Size /Root /Info \\
Skia/PDF&trailer /Size /Root /Info \\
PDFLaTeX&trailer /Root /info /ID /Size\\\hline
\end{tabular}

\end{center}
\caption{Trailer - different trailer keys used by PDF producer tools (all the keys are same across 3 operating systems (Windows, Linux and Mac OS X)).}
\label{tab:trailer}
\end{table*} 

%%%%%%%%%%%%%%%%%%%%%%%%%%%%%%%%%%%%%%%%%%%%%%%%%%%%%%%%%%%%%%%%%%%%%%%
\emph{It is important to notice that we have excluded the elements containing the metadata in the creation of our rules.} Metadata information is stored in a object within the body section of the file. Removing the metadata object or even altering the values present in it has no effect on the functioning of PDF files.

We used the different patterns observed for each section of the PDF file and exploited them to detect the PDF producer tools. We have used regular expressions and expressed them using YARA\footnote{\url{https://github.com/virustotal/yara}} rules.  Figure~\ref{yara} shows an example of a YARA rule written to match one of the text pattern generated by the \texttt{Microsoft Office Word} tool. We observed that, this object is present in every PDF file created using  \texttt{Microsoft Office Word} tool.

%------------------------------------------------------------------------%
 \begin{table*}
  \begin{center}

    \begin{tabular}{|c|l|}
    \hline
      \textbf{PDF section} & \textbf{YARA rule for MicroSoft Office Word}\\\hline
      \multirow{4}{4em}{\textit{Body}} &\textit{string:}\\
      &\textit{\$rule= /4 0 obj\textbackslash{}r\textbackslash{}n\textless{}\textless{}\textbackslash{}/Filter\textbackslash{}/FlateDecode\textbackslash{}/Length [0-9]*\textgreater{}\textgreater{}\textbackslash{}r\textbackslash{}nstream\textbackslash{}r\textbackslash{}n/}\\
      &\textit{condition:}\\
      &\textit{\$rule1}\\
      \hline
   \end{tabular}
    \caption{Example of one YARA rule for an object present in the body part of a PDF file created using Microsoft Office Word tool.}
    \label{yara}
  \end{center}
\end{table*}
%------------------------------------------------------------------------%

We have seen earlier in this section that, for the same content of PDF files, number of objects and the way objects are created differs. Since the coding style is different across tools, the number of YARA rules are also different across tools. Table~\ref{rules} shows the number of rules we have written for  each producer tool. %For some \LaTeX{} chain of tools, we observed that few of our rules can detect the OS. 

%Since, sometimes patterns can be used to detect the OS or the \LaTeX distribution. 

%%%%%%%%%%%%%%%%%%%%%%%%%%%%%%%%%%%%%%%%%%%%%%%%%%%%%%%%%%%%%%%%%%%%%%
\begin{table}
\begin{center}
\begin{tabular}{|l|l|}
\hline
Producer tool & \# rules \\\hline\hline

Acrobat Distiller 		&  13\\
Microsoft Office Word   & 16\\
LibreOffice 			&   15	\\
Ghostscript 			& 15\\
Mac OS X Quartz 		& 30\\
PdfTeX 					& 31\\
SKia/PDF 				& 12\\
Cairo 					& 16\\
xdviPDFmx 				& 13\\
LuaTeX 		            & 22\\
PDFLaTeX 				& 9\\ 
\hline
%total & 769&&

\end{tabular}
%\caption{Our methodology results for PDF producer tool detection.}
%\label{14}
\caption{Rules for PDF producer tools.}
\label{rules}
\end{center}
\end{table}
%------------------------------------------------------------------------%

\begin{figure} 
\small
\begin{forest}
  [pdfTeX
    [Common rules
       [\textit{18}]
    ] %done with left hand graph
    [OS Specific rules
      [Windows
        [\textsc{3}
        ]
      ]
      [Mac OS
        [\textsc{4}
        ]
      ]
      [Linux
          [Linux+Windows
            [\textit{1}]
          ]
          [Linux+Mac OS
              [\textit{3}]
          ]
          [Linux
              [\textit{2}]
          ]
        ]
      ]
    ]
\end{forest}
\caption{Number of YARA rules across three Operating Systems for \textit{pdfTeX} tool.}
\label{fig:tree}
\end{figure}
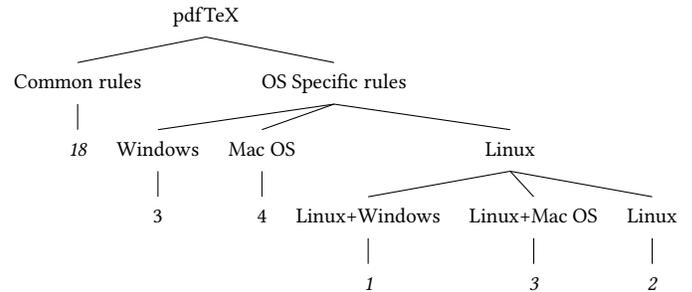

In the header section, we have already provided an example that helps in the detection of OS and \LaTeX{} distribution used (LuaTeX tool). During our analysis of 11 producer tools, for some \LaTeX{} chain of tools, we observed that few of our rules can detect the OS. Figure~\ref{fig:tree} provides an example of the tool \texttt{pdfTeX} and number of rules associated across different Operating Systems. In the Figure~\ref{fig:tree}, we can observe that \texttt{pdfTeX} has 18 common rules for three OSs and some rules are specific to one or two OSs. We used these OS specific rules in the detection of the OS used to create the PDF files.

%\end{document}

% 
%\begin{center}
%\lstset{ %
%    basicstyle=\ttfamily\footnotesize,
%    frame=single,
%    keywordstyle=\color{blue},
%    language=Bash,
%    showstringspaces=false,
%    caption={MikTeX distribution pdfTeX tool on Linux Operating System.},
%    captionpos=b,
%    label=pdftex1,
%    morekeywords={blue},
%}
%\begin{lstlisting}[basicstyle=\small]
%11 0 obj
%<<
%/Font << /F8 7 0 R >>
%/XObject << /Im1 1 0 R >>
%/ProcSet [ /PDF /Text /ImageC /ImageI ]
%>>
%\end{lstlisting}
%\end{center}
%  
%\textcolor{red}{to do, I will comw to it once done with other things}

To conclude, the patterns found in different sections of the PDF file can be exploited to detect the PDF producer tools.

\section{Results and observation}\label{sec:res}
In this section we first describe the dataset used to test the accuracy of our methodology of coding style. Then we show the results for the detection of 11 popular PDF producer tools.

\subsection{Dataset}

Conference proceedings and preprint are clearly the best options to obtain large number of PDF files. We have downloaded PDF files from the preprints of scientific community. Our first dataset was downloaded from the Cryptology ePrint Archive\footnote{\url{https://eprint.iacr.org/}}, it includes 11405 PDF documents from 2004 to 2018. We observed that all the PDF documents were originally compiled by the authors using a PDF producer tool of their choice.

Open archive HAL\footnote{\url{https://hal.archives-ouvertes.fr/}} administrators provided us with an access to a second set of dataset of PDF files. It consists of 544460 PDF files from 1996 to 2019. HAL provides options to either directly submit a PDF file or submit source files and create a PDF file using HAL's pdf producer tool \texttt{PdfLaTeX}. We observed that around 47\% of the PDF files in HAL dataset were compiled by HAL while the rest 53\% were originally compiled by the author(s) using a PDF producer tool of there choice.  All the PDF files submitted to HAL are concatenated with a cover page using Apache PDFBOX. Hence all the PDF files published are modified. \\

Our datasets are respectively called IACR dataset and HAL dataset throughout the paper for convenience. The IACR and HAL PDF files are not anonymous, i.e. they include the name of the author(s). It is also possible to find a description (metadata) of author names, title, year of publication of each PDF file on there respective websites. We observed that the metadata of 53\% of the PDF files in HAL dataset are likely to be original values of authors PDF producer tools. For the rest of the HAL's PDF files, we have clues that metadata were altered by HAL and still we have considered all these PDF files in HAL dataset for our experiments.

We have tested our rules on the PDF files of two dataset: IACR and HAL. We applied our rules to detect the PDF producer tool and then the results obtained using our tool are validated using the PDF producer tool name found in the metadata field producer. \emph{We would like to clarify that, during the PDF producer tool detection, we have not considered the metadata object present in the PDF file. Our rules are used on the rest of the PDF file. Metadata field producer was used only to validate our results.}

\textbf{Cleaning the dataset}:  Then based on the values found in metadata field producer, IACR dataset includes 10892 (96\%) PDF files from the 11 tools we are inspecting and 497944 (91\%) from HAL dataset. We applied our rules on these sets of PDF files and findings are described below. Table~\ref{32} provides the total number of PDF files associated to each tool that we have examined. It also shows PDF files created using rare producer tools, these rare tools are out of scope of this paper. 
  
\begin{table}
\renewcommand{\arraystretch}{1.2}
%\begin{center}
\begin{tabular}{|l|l|l| }
\hline
Producer tool &IACR&HAL\\\hline\hline

Ghostscript 		&	 1200 (11\%) & 115948 (21\%)\\
Acrobat Distiller 	& 369 (3\%) & 96157 (18\%)\\
Microsoft Office Word& 147(1.3\%) &15995 (2\%)\\
Mac OS X Quartz 	& 62 (0.5\%) & 7767 (1.4\%)\\
LibreOffice 		& 4  &451 (0.08\%)\\
SKia/PDF 			 & 0  & 88 (0.01\%)\\
LuaTeX 				&17  &67 (0.01\%)\\
PDFLaTeX 			& 1  & 249147 (46\%)\\
xdviPDFmx 			& 1347 (12\%) & 836 (0.2\%)\\
Cairo 			    & 0  & 1100 (0.2\%)\\
PdfTeX 				& 7745 (68\%) & 10375 (2\%)\\
Rare tools 			& 513 (4\%) & 46529 (9\%)\\
\hline
\end{tabular}
%\caption{PDF producer tools and PDF associated to each tool in our dataset.}
%\label{32}
%\end{center}
\caption{PDF producer tools and number of PDF files  associated to each tool in our dataset.}
\label{32}
\end{table}
%%%%%%%%%%%%%%%%%%%%%%%%%%%%%%%%%%%%%%%%%%%%%%%%%%%%%%%%%%%%%%%%%%%%%%%

\subsection{Detection of 11 PDF producer tools}
We first attempted to use the patterns found in each section separately and the results are given in Table~\ref{stats}. It shows when the producer detected by our tool is correct, wrong or when it is unable to detect a producer tool. Table~\ref{stats} shows that detection of the producer tool based only on the header and xref table is not efficient for IACR dataset. The results for these two sections are improved for HAL dataset. Body and trailer sections offer better perspective for both the dataset. However, the results are not conclusive.   

%%%%%%%%%%%%%%%%%%%%%%%%%%%%%%%%%%%%%%%%%%%%%%%%%%%%%%%%%%%%%%%%%%%%%%%
\begin{table}
\renewcommand{\arraystretch}{1.2}
\begin{center}
\small
\begin{tabular}{|c|l|l|l|l|}

%%%%%%%%%%%%%%%%%%%%%%%%%%%%%%%%%%%%%%%%%%%%%%%%%%%%%%%%%%%%%%%%%%%%%%%
\multicolumn{5}{c}{\textbf{IACR (10892 PDF files)}}\\ \hline
Detection &  Header & Body & Xref Table & Trailer \\ \hline\hline
Correct & 1676 (15\%)&7302 (67\%)&898 (8\%)&8641 (79\%)\\
Wrong &8325 (76\%)	&1839 (17\%)&1277 (12\%)&141 (1\%)\\
No result & 891 (8\%)& 1751 (16\%) & 8717 (80\%) & 2110 (19\%)\\
%%%%%%%%%%%%%%%%%%%%%%%%%%%%%%%%%%%%%%%%%%%%%%%%%%%%%%%%%%%%%%%%%%%%%%%%%%%%
\hline
\multicolumn{5}{c}{\textbf{HAL (497944 PDF files)}}\\ \hline
Detection &  Header & Body & Xref Table& Trailer \\ \hline\hline
Correct&235859 (47\%)&197069 (40\%)&232540 (47\%)&229060 (46\%)\\

 Wrong & 204214(41\%) & 241101 (48\%) & 232104 (47\%) &  259988 (52\%)\\
 No result & 57871 (12\%)  & 59774  (12\%)  & 33300 (7\%)&8896 (2\%)\\

\hline
\end{tabular}
\end{center}
\caption{Detection of PDF producer tools for header, body, xref and trailer section.} %(Correct, wrong predictions, prediction for 1 tool and among 2 tools).}
\label{stats}
\end{table}
%%%%%%%%%%%%%%%%%%%%%%%%%%%%%%%%%%%%%%%%%%%%%%%%%%%%%%%%%%%%%%%%%%%%%%

For each individual section, our tool can sometimes detect two producer tools. This case is considered as a wrong prediction in Table~\ref{stats}. We have evaluated the frequency of this event in Table~\ref{stats2tool}. Two cases are possible. The detection can be \emph{confused}: for the same PDF file, our tool detects the correct producer and an another one (incorrect). We have an error when the two tools detected are incorrect.

The results in Table~\ref{stats2tool} shows that the detection based on single section of the PDF file has too much uncertainty. For instance the header section for IACR dataset resulted in 66\% of confused detection, since pdfTeX and LuaTeX use same producer magic number. Our tool often detects 2 tools for PDF files produced by either pdfTeX and LuaTeX.   To improve our tool, we have combined the results of each section using a majority vote. In case of equality, \textit{i.e.} two producers have received two votes, we have considered that our tool takes a wrong decision. 

%For instance, if 3 sections detect correct tool and 1 detects incorrect and so on, using majority voting this detection is considered correct.  
 
%%%%%%%%%%%%%%%%%%%%%%%%%%%%%%%%%%%%%%%%%%%%%%%%%%%%%%%%%%%%%%%%%%%%%%%
\begin{table}[!ht]
\renewcommand{\arraystretch}{1.2}
\begin{center}

\small
\begin{tabular}{|c|l|l|l|l|}

%%%%%%%%%%%%%%%%%%%%%%%%%%%%%%%%%%%%%%%%%%%%%%%%%%%%%%%%%%%%%%%%%%%%%
\multicolumn{5}{c}{\textbf{IACR (10892 PDF files)}}\\ \hline
Detection &  Header & Body & Xref Table & Trailer \\ \hline\hline
Confused&7213 (66\%) & 433 (4\%) & 1228 (11\%) & 130 (1\%)\\
Error& 1112 (10\%)	 &  1406 (13\%)   &  49 (0.4\%) &  11 (0.1\%)	\\
%No detection &&&&\\
%%%%%%%%%%%%%%%%%%%%%%%%%%%%%%%%%%%%%%%%%%%%%%%%%%%%%%%%%%%%%%%%%%%%%%%%%%%%
\hline
\multicolumn{5}{c}{\textbf{HAL (497944 PDF files)}}\\ \hline
Detection &  Header & Body & Xref Table& Trailer \\ \hline\hline

%Confused & 38198(8\%) &  15023 (3\%) &  17272 (3.4\%) &  8743 (2\%)\\
Confused & 19673 (4\%) & 44751 (9\%)&  16028 (3.2\%) &   153 (0.03\%)\\
Error& 166016 (33\%) & 226078 (45\%)& 216076 (43\%) &  251245 (50\%)\\
%No detection &&&&\\

\hline
\end{tabular}
\end{center}
\caption{Frequency of detection of 2 PDF producer tools.} %(Correct, wrong predictions, prediction for 1 tool and among 2 tools).}
\label{stats2tool}
\end{table}
%%%%%%%%%%%%%%%%%%%%%%%%%%%%%%%%%%%%%%%%%%%%%%%%%%%%%%%%%%%%%%%%%%%%%%%

Table~\ref{tab:final1} shows results for the detection of producer tools for combination of all the sections using majority votes. Our tool finds the correct PDF producer tool 74\% of the time for IACR dataset and 48\% for HAL dataset. PDF files in HAL dataset are modified using PDFLaTeX tool. These modified PDF files includes coding style of both PDFLaTeX and the original tool initially used to create the PDF file. Currently our tool does not apply to the detection of modified/concatenated PDF files and hence the results obtained for HAL dataset are less impressive.

%%%%%%%%%%%%%%%%%%%%%%%%%%%%%%%%%%%%%%%%%%%%%%%%%%%%%%%%%%%%%%%%%%%%%%%%%%%%%%%%%
\begin{table}[!ht]
\renewcommand{\arraystretch}{1.2}
\begin{center}
\begin{tabular}{|c|l|l|}
 \hline

\textbf{Detection}  & \textbf{IACR} (10892) & \textbf{HAL} (497944)\\\hline
producer tool &8018 (74\%) & 239967  (48\%)\\
OS &    3344  (29\%) & 105930  (19\%)\\ 

\hline
\end{tabular}
\end{center}
\caption{Detection of PDF producer tool and OS using combination of different sections.}
\label{tab:final1}
\end{table}

%%%%%%%%%%%%%%%%%%%%%%%%%%%%%%%%%%%%%%%%%%%%%%%%%%%%%%%%%%%%%%%%%%%%
\paragraph{OS Detection:} Along with detection of the PDF producer tool used, coding style can also reveal the OS. Table~\ref{tab:final1} shows corresponding results obtained for the detection of OS. Header, trailer and xref section's coding style does not vary much across different Operating Systems and hence predicting the OS using these sections is harder and sometimes impossible. But using the body section's coding style it is possible to detect the OS for some producer tools. During the analysis of coding style we observed that body section coding style includes one or two objects that can be used to detect the Operating System. For instance, pdfTeX (MikTeX distribution) uses some keys across Microsoft Windows, Linux and Mac OS that are distinguishable from one another and leads to the detection of OS. Even though we could identify OS for fewer number of PDF files (Table~\ref{tab:final1}), these results are not conclusive for the majority of the files in our dataset.

\subsection{Results for each producer tool}

Our tool is more efficient for the detection of certain producer tools. Table~\ref{14} shows the differences between the different tools evaluated. The detection is still done using a majority vote over all the parts of the PDF file. PDF files produced by \texttt{Microsoft Office Word, Mac OS X Quartz, PDFLaTeX} and \texttt{LibreOffice} are detected with the higher accuracy (more than 90\%).

As previously explained the PDF files in HAL dataset are modified, some parts of the PDF files are replaced by the coding style of PDFLaTeX tool. Since we use the majority votes, the results for tools other than PDFLaTeX (91\%) lead to wrong detection. If the PDF file is not modified, our rules detects the PDF producer tool with higher accuracy and results obtained for detection of producer tools in IACR dataset supports it.

%%%%%%%%%%%%%%%%%%%%%%%%%%%%%%%%%%%%%%%%%%%%%%%%%%%%%%%%%%%%%%%%%%%%%%%%%%%%%%%%%%%%%%%
\begin{table*}[!ht]
\renewcommand{\arraystretch}{1.2}
\begin{center}
\begin{tabular}{|l|l|l|l|l|l|l|}
\hline
Producer tool  & \# PDF IACR &\# detection & \# PDF HAL& \# detection\\
\hline\hline
Acrobat Distiller 	& 369  &  269   (73\%) & 96157 	& 3497  (4\%)	\\
Microsoft Office Word  & 147  &  141	(96\%) & 15995 	& 4388  (28\%)	\\
LibreOffice 	& 4    &   4    (100\%)& 464 	& 451    (95\%)	\\
Ghostscript 	 & 1200 &  993  (83\%) & 115948 & 1940  (2\%)	\\
Mac OS X Quartz  & 62   &  61    (98\%) & 7767 	& 2111  (27\%)	\\

SKia/PDF 	 & 0    &   0    	 & 88	 	& 67   (76\%)	\\
Cairo 	 & 0    &   0    	 & 1100	& 357    (32\%)	\\
PdfTeX 		 & 7745 &  6001  (77\%) & 10375	& 743   (7\%)	\\
xdviPDFmx 	 & 1347 &  536  (40\%) & 836 	& 102   (12\%)	\\
LuaTeX 		& 17    &  12    (71\%) & 67 	& 4     (6\%)	\\
PDFLaTeX 	 & 1    &   1    (100\%)& 249147	& 226281 (91\%)	\\

\hline
\end{tabular}
%\caption{Our methodology results for PDF producer tool detection.}
%\label{14}
\caption{Detection of individual PDF producer tool.}
\label{14}
\end{center}
\end{table*}

The results obtained in this section shows that it is possible to detect the PDF producer tool of a PDF file using the coding style. Therefore, creating PDF files without metadata or altering metdata is not enough to hide information.

\subsection{Consistency of  online PDF tools}
%We have created PDF files using online PDF compressors and producer tools that are freely available.
 We have created PDF files using  online PDF tools (29 compressors and 22 producer tools) that are freely available. Then, we have applied our detection tool on the obtained PDF files and we have compared the producer detected to the value of the producer in the metadata. Table~\ref{tab:online_creator} and~\ref{tab:online_optimizer} shows the list of online tools we have evaluated and the results we have observed. We found some inconsistencies for 6 PDF compressor tools (in red in Table~\ref{tab:online_optimizer}). A producer is advertised in the metadata but it is actually anthoer software that has been used.
 
  It should noted that some of the online tools advertise that they use their own PDF producer software but it is actually a generic software that has been used. It means that most of the time a user can install locally the PDF creation tool instead of uploading his/her sensitive PDF files on an online service which has questionable trust. 

\begin{table}[ht]
\renewcommand{\arraystretch}{1.2}
\small
\begin{center}
\begin{tabular}{|l|r|r|}
\hline
\textbf{Online tool} & \textbf{Producer name in Metadata} & \textbf{Tool detected}\\ \hline\hline
pdf.io	& LibreOffice 6.0 & LibreOffice	 \\%\checkmark \\
Hipdf & Microsoft Word 2013 &	Microsoft Office Word\\
PDFyeah	& LibreOffice 6.1 & LibreOffice \\
google docs & Skia/PDF m76 & qpdf	\\
pdfconvertonline & LibreOffice 5.4 & LibreOffice \\
toPDF & LibreOffice 4.2	& LibreOffice \\
small pdf & Microsoft Word for Office 365 & Microsoft Office Word \\
jinapdf	& Microsoft Word 2016 &	Microsoft Office Word	\\
PDF24 Tools & LibreOffice 6.1& 	 LibreOffice \\
pdf2go& LibreOffice 6.2 & LibreOffice	\\

lightpdf & lightpdf.com... & No match \\
altocompress &	None & Microsoft Office Word  \\
pdfaid.com & pdfaid using ABCpdf... &	No match \\
doc2pub & Neevia PDFcompress... & No match \\
VeryPDF & http://www.verypdf.com & No match	\\
Sejda & Apache FOP Version 2.3 & No match \\

online2pdf & Online2PDF.com & No match\\
I love pdf & www.ilovepdf.com & No match \\
Free PDF Editor & FreePDF.net PDFill  & No match  \\
PDFcandy & PDF Candy & No match	\\
ZonePDF	& zonepdf.com &	No match \\
sodapdfonline & Soda PDF Online & No match \\

\hline
\end{tabular}
\caption{Analysis of online PDF creator tools.}
\label{tab:online_creator}
\end{center}
\end{table}
%%%%%%%%%%%%%%%%%%%%%%%%%%%%%%%%%%%%%%%%%%%%%%%%%%%%%%%%%%%%%%%%%%%%%%%%%%%%
%%%%%%%%%%%%%%%%%%%%%%%%%%%%%%%%%%%%%%%%%%%%%%%%%%%%%%%%%%%%%%%%%%%%%%%%%%%%
\begin{table}
\renewcommand{\arraystretch}{1.2}
\small
\begin{center}
\begin{tabular}{|l|r|r|}
\hline
\textbf{Online tool}	& \textbf{Producer name in Metadata} &	\textbf{Tool detected}	\\\hline

\textcolor{red}{altocompress} & \textcolor{red}{PDFfiller} & \textcolor{red}{Ghostscript}\\
\textcolor{red}{compress/zipfile} & \textcolor{red}{Same as original file} & \textcolor{red}{pdfTeX}\\
\textcolor{red}{VeryPDF}	& \textcolor{red}{VeryPDF} & \textcolor{red}{Ghostscript}\\
\textcolor{red}{pdfcompressor}&	\textcolor{red}{3-Heights(TM)...}& \textcolor{red}{Acrobat Distiller}\\
\textcolor{red}{small pdf} & \textcolor{red}{3-Heights(TM)}& \textcolor{red}{Acrobat Distiller}\\
jinapdf	& GPL Ghostscript 9.26 & Ghostscript\\
PDFill & GPL Ghostscript 9.23 & Ghostscript\\
pdfzipper & GPL Ghostscript 9.21 & Ghostscript\\
pdf2go	& GPL Ghostscript 9.26 & Ghostscript	\\
PDFcandy & GPL Ghostscript 9.10 & Ghostscript	\\
p2w compresspdf & GPL Ghostscript 9.22	& Ghostscript	\\
PDFyeah	& GPL Ghostscript 9.26 & Ghostscript \\
youcompress &	GPL Ghostscript 9.26 & Ghostscript \\
\textcolor{red}{wecompress} & \textcolor{red}{Same as original file} & \textcolor{red}{Acrobat Distiller}	\\
pdfconvertonline & Aspose.Pdf for .NET 17.1.0 & No match \\
lightpdf & lightpdf.com & No match \\
pdfaid & pdfaid using ABCpdf & No match	\\
PDF resizer & itext-paulo-155... & No match \\
Neevia PDFcompress & Neevia PDFcompress... & No match \\
Sejda	& SAMBox 1.1.50... & No match	\\
PDF-online.com&	3-Heights(TM)...& No match	\\
online2pdf & Online2PDF.com &	No match \\
I love pdf & www.ilovepdf.com & No match \\
ZonePDF	& Aspose.PDF... & No match \\
PDF24 Tools & GPL Ghostscript 9.26	&No match\\
image resize & GPL Ghostscript 9.23 & No match\\
foxit &	Same as original file & No match \\
sodapdfonline & Same as original file & No match\\
Hipdf	& Same as original file & No match \\
\hline
\end{tabular}
\end{center}
\caption{Analysis of online PDF compressor tools.}
\label{tab:online_optimizer}
\end{table}
\section{Related works}\label{sec:related}

Many works have been devoted to PDF malwares detection~\cite{Castiglione2010,Carmony2016,Markwood2017,Maiorca2019a,Maiorca2019b,Chen2020} and to privacy issues related to hidden data~\cite{Aura2006,Garfinkel2014,Mendelman2018,Feng2018}. \\

ArXiv hosts around 1.6 million e-prints in different science fields. ArXiv administrators (\url{https://arxiv.org}) have designed a tool. All the submissions are controlled to check that the material provided is appropriate, topical and meets arXiv's guidelines. \LaTeX{}, AMSLaTeX, PDFLaTeX sources, PDF and HTML with JPEG/PNG/GIF images formats are accepted by arXiv. ArXiv accepts only PDF files that are produced by Microsoft Word compatible software. PDF files created using \LaTeX{} software are not accepted. The sources must be submitted and arXiv's server produces directly the PDF file from the source. Using a guess and determine approach, we were able to reverse-engineer arXiv PDF detection tool. It appears that arXiv detector first tests if the metadata of the PDF file submitted match those of a \LaTeX{} file. If this test is not conclusive, it checks if \LaTeX{} fonts have been used. If both tests failed, arXiv detector considers that the file has been produced by Microsoft Word compatible software. We have created a PDF file using \LaTeX{} software and using other fonts (Listing~\ref{listing1}) and it was accepted by ArXiv. \\

%------------------------------------------------------------------------------%
So far we have found that, only ArXiv detector is very similar to our tool. However, it only detects if a PDF file has been produced using \LaTeX{} software using metadata and \LaTeX{} fonts. Whereas, our tool can identify 11 PDF producers by analyzing the PDF coding style even when the metadata have been removed and PDF file is sanitized. \\

Our PDF producer detector shares similar idea with BinComp~\cite{Rahimian2015}, a tool designed to detect which compiler has been used to create a binary. This tool has been designed to perform code authorship attribution. The overall goal of such a software is to help to  identify the authors of malicious software. This domain has been very active in the last years~\cite{Ferreira2017,Simko2018,Kalgutkar2019}. Our tool is designed to identify coding style pattern used by PDF producer tools to detect PDF producer tool.

\begin{center}
\lstset{ %
    basicstyle=\ttfamily\footnotesize,
    frame=single,
    keywordstyle=\color{blue},
    language=Bash,
    showstringspaces=false,
    caption={PDF compiled using \texttt{pdfLaTeX} With MS Word-style font.},
    captionpos=b,
    label=listing1,
    morekeywords={blue},
}
\begin{minipage}{3.5in}
\begin{lstlisting}[basicstyle=\small]
\documentclass{article}
\usepackage[T1]{fontenc}
\usepackage{newtxmath,newtxtext}
\usepackage{lipsum}
% just to generate dummy text
\begin{document}
\lipsum
\end{document}
\end{lstlisting}
\end{minipage}
\end{center}

\section{Conclusion}

Coding style can be exploited to identify which software has been used to create a PDF file. The results obtained with our tool shows that it is working with high accuracy. It is more robust than just looking at the metadata fields of the file, which is highly unreliable (it can be altered or removed from the file). In our work, we exploit patterns in the four sections of PDF files. It is more difficult for an adversary to manipulate the content of a PDF file to fool our tool.  \\

An interesting question is that is it possible to extend our tool to other PDF producers? In other words, is it possible to automatize the creation the Yara rules to identify a producer tool? An exhaustive approach would consist in taking a string of fixed length in a PDF file. This string can be included in a regular expression and the new rule can be tested to check if it is accurate. If it is not, the strings can be extended until it is accurate. Unfortunately, this approach is time consuming: the complexity depends on the file size. Applying machine learning techniques could be a promising future work. 

\section{Acknowledgments}
This work has been supported by the SIDES 3.0 project (ANR-16-DUNE-0002) funded by the French Program Investissement d'Avenir and the Grenoble Alpes Cybersecurity Institute CYBER@ALPS under contract ANR-15-IDEX-02. \\
\bibliographystyle{ACM-Reference-Format}
\bibliography{ref}

\end{document}